\documentclass[showpacs,twocolumn,aps,prl,groupedaddress]{revtex4}
\newcommand{\bra}[1]{\langle #1|}
\newcommand{\ket}[1]{|#1\rangle}

\usepackage{graphicx}
\begin{document}
\title{Decoherence in a Josephson junction qubit}
\author{A. J. Berkley}
\email[]{berkley@physics.umd.edu}
\author{H. Xu}
\author{M. A. Gubrud}
\author{R. C. Ramos}
\author{J. R. Anderson}
\author{C. J. Lobb}
\author{F. C. Wellstood}
\affiliation{Center for Superconductivity Research, Department of Physics, University of Maryland, College Park, MD 20742}
\date{\today}
\begin{abstract}
\setlength{\parindent}{0.5in} 
The zero-voltage state of a Josephson junction biased with constant current
consists of a set of metastable quantum energy levels.  We probe the spacings
of these levels by using microwave spectroscopy to enhance the escape rate to
the voltage state.  The widths of the resonances give a measurement of the
coherence time of the two metastable states involved in the transitions.  We
observe a decoherence time shorter than that expected from dissipation alone in
resonantly isolated 20x5 $(\rm \mu m)^2$ area Al/AlOx/Al junctions at 60 mK.
The data is well fit by a model that includes the dephasing effects of both
low-frequency current noise and the escape rate to the voltage
state.  We discuss implications for quantum computation using current-biased
Josephson junction qubits, including limits on the minimum number of levels needed in the
well to obtain an acceptable error limit per gate.

\end{abstract}
\maketitle
\setlength{\parindent}{0.5in}  
Research in the 1980s definitively showed that the phase difference across a
single current-biased Josephson junction can behave quantum-mechanically
\cite{webb,martiniselq}.  The recent proposal that an isolated current-biased
Josephson junction could serve as a qubit \cite{ramos} in a quantum computer
has preceeded a resurgence of interest in this simple system
\cite{martinisrabi,hanrabi,blais,strauch}.

Designing a quantum computer based on isolated Josephson junctions raises many
issues.  Isolation of the junction from its bias leads must be achieved by
controlling the high frequency electromagnetic environment that the junction
couples to \cite{martiniselq}.  At the very least, this isolation must be effective around the resonant
frequency of the junction.  In addition, at lower frequencies, current noise will tend to cause decoherence in the junction state.
Also, during typical gate operations the junction must operate
in a strongly anharmonic regime that can be reached by applying a large bias
current through the junction.  In this high bias regime however, there is an
increased escape rate from the upper qubit state.  In this Letter, we describe
how both the escape rate and low frequency current noise cause decoherence
and report results on measurements of these effects in Al/AlOx/Al
Josephson junctions.

\begin{figure}[b]
\includegraphics[width=2.2in,height=1.5in]{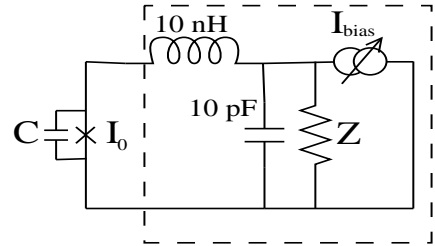}
\caption{Circuit schematic of current-biased Josephson junction.  All the elements
in the dashed box are represented by an equivalent resistance, $R(\omega)$.
}\label{fig:isolation}
\end{figure}
Consider a
Josephson junction shunted by capacitance $C$, having a critical current $I_0$,
and a parallel shunting impedance $R(\omega)$ due to the external wiring (see
Fig.  \ref{fig:isolation}).  The supercurrent $I$ through the junction is given
by the Josephson relation $I = I_0 \sin(\gamma)$, and the voltage by $V =
(\Phi_0/2 \pi) \rm d\gamma/\rm{dt}$, where $\gamma$ is the gauge-invariant
phase difference between the superconducting wavefunctions on each side of the
junction.  For $I<I_0$, the phase $\gamma$ may be trapped in a metastable
well of the Josephson washboard potential, $U = -(\Phi_0/2\pi)I_0 \cos \gamma -
(\Phi_0/2\pi) I \gamma$, or it may be in a running state with a non-zero
average dc voltage \cite{tinkham}.  

\begin{figure}[b]
\includegraphics[width=2.2in,height=1.5in]{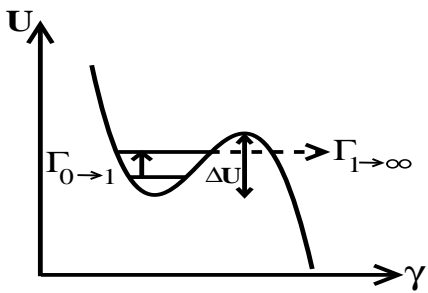}
\caption{Josephson junction potential energy $U$ as a function of the phase
difference $\gamma$.} \label{fig:well} \end{figure}

Quantizing the single junction system in the absence of dissipation leads to
metastable states that are localized in the wells (see Fig. \ref{fig:well}) and
adds the possibility of escape to the continuum running states by quantum tunneling from the
$i$th level with a rate $\Gamma_{i\rightarrow\infty}$.  The energy barrier
$\Delta U= (I_0 \Phi_0/\pi)\cdot(\sqrt{1-(I/I_0)^2} - (I/I_0) {\rm acos}(I/I_0))$
to the continuum decreases as the bias current is increased, leading to a rapid
increase in the tunneling rate with bias current \cite{alvarez}:
\begin{equation}\label{eqn:tunnrates}
\Gamma_{i\rightarrow\infty} = \omega_p \frac{\left(432 N_s\right)^{i+1/2}}{\left(2\pi\right)^{1/2} i!} e^{-36 N_s / 5}
\end{equation}
where $\omega_p = \sqrt{\frac{2\pi I_0}{\Phi_0 C}} \left( 1 - \left(I /
I_0\right)^2\right)^{1/4}$ is the classical oscillation frequency and $N_s = \Delta U / \hbar \omega_p$ is
approximately the number of levels in the well.
As the energy barrier is lowered, the energy of the states in the well move closer
together and the well becomes more anharmonic until, at $I = I_0$, the energy
barrier disappears.

\begin{figure}[b]%
\includegraphics[width=3in,height=2.5in]{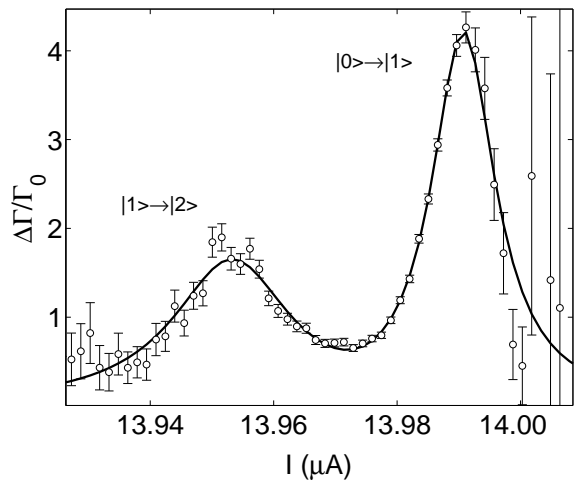}
\caption{Enhancement of escape rate under 5.7 GHz microwave drive.  Left axis is the difference in escape rate with and without
microwaves divided by the escape rate without microwaves.  The large error
bars on the left and right of figure come from a lack of counts in the
escape histogram.  The right peak is $\ket{0}\rightarrow\ket{1}$ quantum transition,
while left peak is $\ket{1}\rightarrow\ket{2}$.  
Solid line is a Lorentzian fit to two peaks.}\label{fig:twopeaks}
\end{figure}

The observed escape rate of the system from the zero-voltage state to the
finite voltage state at a given bias point is $\Gamma = \Sigma_{i=0}^n
\Gamma_{i\rightarrow\infty} P_i$, where $P_i$ is the probability of the
junction being in the $i$th state.  An ac current, $I_{ac}$ (either external or
thermally generated) can induce transitions between levels $i$ and $j$ in the
well with a rate $\Gamma_{i\rightarrow j} \propto \left|\frac{\Phi_0}{2 \pi}
I_{ac} \bra{i}\gamma \ket{j}\right|^2$.  Since $\Gamma_{1\rightarrow\infty}
\simeq 500 \Gamma_{0\rightarrow\infty}$ for typical junction parameters, one
expects to see a large enhancement in the escape rate if a microwave
source is used to resonantly excite the system from the ground state $\ket{0}$ to the
first excited state $\ket{1}$ (see Fig. \ref{fig:twopeaks}).\cite{martiniselq}

Each microwave resonance in this system will be broadened due to the
interaction of the junction with noise transmitted via the wiring and described
by the interaction Hamiltonian $H_{int} = - (\Phi_0 / 2 \pi) I_{noise} \gamma$.
Thermal noise and dissipation at the transition frequencies will cause changes
in the populations of the states.  At low frequencies, the resonant terms are
insignificant and the noise only causes dephasing.

Considering just the ground state $\ket{0}$ and the first excited state
$\ket{1}$, transitions arise from thermal excitation from $\ket{0}$ to
$\ket{1}$, a $1/R C$ decay rate from $\ket{1}$ to $\ket{0}$, and tunneling to
the continuum,
$\Gamma_{i\rightarrow\infty}$ for $i=0$ and $1$.  At temperature $T$, the
combined thermal and dissipative transition rates are:\cite{chow}
\begin{equation} \Gamma_{0\rightarrow1} = \frac{1}{R C (\exp(\Delta E / k T) -
1)} \end{equation} \begin{equation} \Gamma_{1\rightarrow0} = \frac{1}{R C (1 -
\exp(-\Delta E / k T))} \end{equation}
where $\Delta E = E_1 - E_0$ is the difference in energy between the two levels.  For $k T \ll \Delta E$, the upward thermal transition rate is much smaller than the
downward rate.  From Eqn. \ref{eqn:tunnrates} the tunneling to the
continuum is much smaller for the ground state than the excited state in the
anharmonic region of interest where $\Delta U / \hbar \omega \simeq
3$.\cite{alvarez}  Thus we expect that the spectroscopic width of the $\ket{0}\rightarrow\ket{1}$ transition is
\begin{eqnarray}\label{eqn:specwidth}
\Delta \omega = \Gamma_{1\rightarrow\infty} + \Gamma_{0\rightarrow\infty} + \Gamma_{0\rightarrow1} + \Gamma_{1\rightarrow0} \simeq \Gamma_{1\rightarrow0} + \Gamma_{1\rightarrow\infty}\nonumber \\ \simeq 1 / RC + \Gamma_{1\rightarrow\infty}
\end{eqnarray}
Equations \ref{eqn:tunnrates} and \ref{eqn:specwidth} imply that the level broadening, $\Delta\omega$, depends on bias through the $\Gamma_{1\rightarrow\infty}$
term and should exceed $1/RC$ as the bias current approaches $I_0$.

To understand results on a real junction we must also take into account the
dephasing effects of any current noise in the system.  For sufficiently low
frequencies, we can model this non-resonant decoherence mechanism as a simple
smearing of the response with bias.  This should result in a broadening of the
spectroscopic width that depends on how sensitive the resonant frequency, $\omega$, is to
changes in current, $\partial \omega / \partial I$.  An rms current noise
$\sigma_I$ should produce an additional contribution to the spectroscopic width
of $2 \sigma_I \partial \omega / \partial I$.  Including this current noise
contribution in the previous form for the spectroscopic width gives:
\begin{equation}\label{eqn:withcnoise} \Delta \omega
\simeq 1/RC +\Gamma_{1\rightarrow\infty}+2 \sigma_I \partial \omega / \partial I\end{equation}
Both the second and third terms in Eq. (\ref{eqn:withcnoise}) depend on bias, so that care must be taken in disentangling the two effects.

Using double angle evaporation, we fabricated $20\times5$ $(\mu m)^2$ 
Al/AlOx/Al Josephson junctions with $J_c\simeq14$ $\rm A/cm^2$.  Direct
measurements of the junction current-voltage characteristics showed a subgap
resistance of more than $10^4$ $\Omega$ at 20 mK.  Escape rate measurements were
made in an Oxford Instruments Model 200 dilution refrigerator with a 20 mK base
temperature.  We were able to tune the critical current of the junction by
means of a superconducting magnet.  The junctions were partially isolated from
the bias leads by a 10 nH surface mount series inductor and a 10 pF capacitive
shunt across the dissipative $50$ $\Omega$ transmission line leads (see Fig.
\ref{fig:isolation}).  This isolation scheme was designed so that at the plasma frequency, the effective shunt
resistance due to the leads would be stepped up from $50$ $\Omega$ to much more than $10^3$ $\Omega$, increasing the intrinsic Q of the system.  To perform escape
rate measurements, we start a timer and then ramp the current slowly (5 mA/s)
using an HP 33120A function generator through a $47$  k$\Omega$ resistor and
monitor the junction voltage with a 2SK170 FET followed by an SRS560 amplifier.  This
output voltage is used to trigger the stop of timing, which is handled by a 20
MHz clock.  Escape events were binned in time with width $t_w \simeq$  $50 ns$ to
create a histogram $H(t_i)$.  The escape rate is then $\Gamma(t_j) =
\frac{1}{t_w} \ln\left[\Sigma_{i=j}^\infty H(t_i) / \Sigma_{i=j+1}^\infty
H(t_i)\right]$.  We convert the time axis to current by calibrating the ramp
current as a function of time.

\begin{figure}[b]%
\includegraphics[width=3in,height=2.5in]{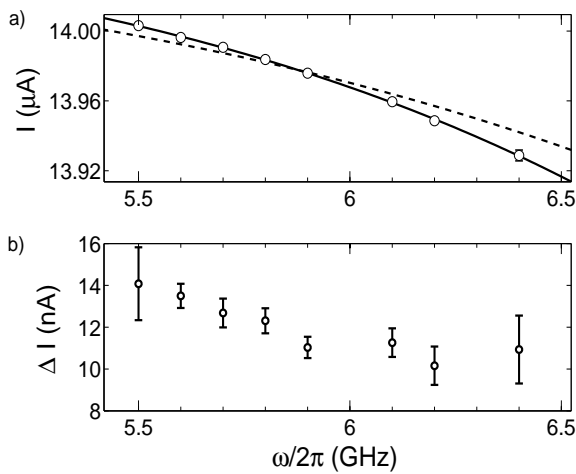}
\caption{(a) Drive frequency vs. center of the $\ket{0}\rightarrow\ket{1}$ resonance peak for $I_0=14.12\mu A$. 
The solid line is a smooth fit used only to extract
a local slope.  The dashed line is a fit to theory.  (b) Full widths of each resonance for $I_0=14.12\mu A$. (Data set \#050902)}\label{fig:wvsi}
\end{figure}

We determine the spacing of the energy levels by comparing escape rate
curves with $(\Gamma_m)$ and without $(\Gamma_0)$ a small microwave drive current applied.
Figure \ref{fig:twopeaks} shows $\Delta \Gamma / \Gamma_0 = (\Gamma_m-\Gamma_0)
/ \Gamma_0$ for a 5.5 GHz microwave signal.  We chose the power so that $\Delta
\Gamma / \Gamma_0 \lesssim 10$ on resonance to ensure the occupancy of
$\ket{1}$ is small.  Two Lorentzian peaks are apparent, corresponding to the
$\ket{0}\rightarrow\ket{1}$ and $\ket{1}\rightarrow\ket{2}$ transitions.  By
measuring $\Delta\Gamma / \Gamma_0$ for different applied microwave
frequency, we can measure how the bias current changes the energy level spacing
 of the $\ket{0}\rightarrow\ket{1}$ transition (see Fig.  \ref{fig:wvsi}a).

The data in Fig. \ref{fig:wvsi}a also allows us to compute $\partial \omega/ \partial I$ and convert 
the full width at half-maximum $\Delta I$ measured at each frequency (see Fig.
\ref{fig:wvsi}b) to a width in frequency, $\Delta \omega$, or
the spectroscopic coherence time associated with the two levels, $\tau = 1 / \Delta \omega$.

\begin{figure}[b]
\includegraphics[width=3in,height=2.5in]{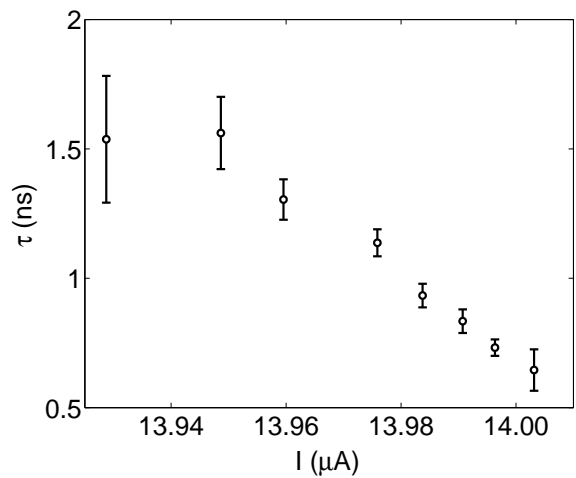}
\caption{Coherence time $\tau$ vs. bias current, I.  Note that the escape rate from the
ground state at 13.93 $\mu$A is around $10^{3}$/s while at 14.01$\mu$A, it is
around $3\times10^{6}$/s.  \label{fig:Qvsi}} \end{figure}

Figure \ref{fig:Qvsi} shows the coherence time $\tau$ as a function of the
center current of each $\ket{0}\rightarrow\ket{1}$ peak.  We note that the
coherence time decreases markedly as $I$ approaches $I_0\simeq14.12\mu A$,
consistent with escape rate limiting of the lifetime of the upper state and
excess low frequency current noise as in Eq. 5.

\begin{figure}[t]
\includegraphics[width=3in,height=2.5in]{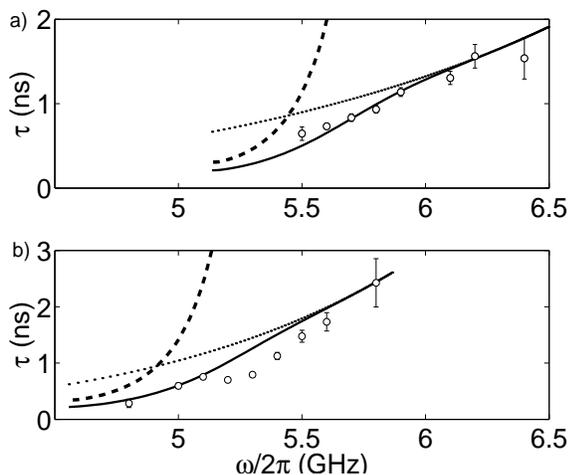}
\caption{Coherence time, $\tau$, vs. bias current I. 
Solid lines are the theoretical
fit for each data set.  Lower frequency corresponds to larger current.  The
parameters for the fit in (a) are $I_0 = 14.12\ \mu A$ and $C = 3.7\ pF$.
For the fit in (b) $I_0 = 10.645\ \mu A$ and $C = 3.7\ pF$.  The dashed lines represent
the contribution from the escape rate, and the dotted lines the contribution from current noise.}\label{fig:Qvsftwo}
\end{figure}

In principle, it is possible for the effective shunting impedance $R(\omega)$
to vary with frequency in such a way as to generate the changes in
$\tau(\omega)$ seen in Fig. \ref{fig:Qvsi}.  We can rule out this explanation
for the overall behavior of $\tau(\omega)$ by changing the critical current of
the junction and remeasuring at the same frequency. Such a process changes
$\Gamma_{i\rightarrow\infty}$ but not $R(\omega)$ in Eq. 5.  Results for two different $I_0$'s
are plotted in Fig.  \ref{fig:Qvsftwo}a and \ref{fig:Qvsftwo}b.  Comparison of
Figs. \ref{fig:Qvsftwo}a and \ref{fig:Qvsftwo}b reveals that the coherence time
at fixed frequency is lower for larger $I_0$.  Since this measurement is at
fixed frequency, the effect can not be due to $R$ varying with frequency.

To distinguish the effects of current noise and escape-rate broadening in Eq.
\ref{eqn:withcnoise}, we need to obtain an independent measure of the junction
parameters.  For the low critical current data, we fit the escape rate curves
without microwaves \cite{bhl} and find $I_0 = 10.65 \pm 0.01\ \mu A$, $C =
3.7 \pm 0.3\ pF$, and $T = 60 \pm 3\ mK$.  The 60 mK temperature was 40 mK
above the base temperature, probably due to self-heating.  We also numerically
solved Schr\"odinger's equation (with hard wall boundary conditions) and chose
$I_0$ and $C$ to fit the data in Fig.  \ref{fig:wvsi}a (dashed line).  This
yielded $I_0 = 10.645 \pm 0.01\ \mu A$ and $C = 3.7 \pm 0.1\ pF$. The same
analysis for the high $I_0$ case gives $I_0 = 14.12 \pm 0.01\ \mu A$ and
$C = 3.7 \pm 0.1\ pF$.  The quantitative disagreement in Fig.  \ref{fig:wvsi}
may come from not including corrections to the center peak locations due to
current noise \cite{asc}, the energy level shifting due to damping, or a
frequency dependent impedance (such as is suggested at 5.2 GHz in Fig.
\ref{fig:Qvsftwo}b).

We now fit the coherence time data in Fig. \ref{fig:Qvsftwo} by varying $I_0$
and $C$ and comparing the results with the previously determined parameters.
We find $\Gamma_{1\rightarrow\infty}$ by solving Schr\"odinger's equation
numerically.  To estimate the rms current noise, $\sigma_I$, we note that the
full current width at half maximum shown in Fig.  \ref{fig:wvsi}b never
drops below 10 nA.  We thus assign $\sigma_I\simeq5$~nA.  To get a unique fit, we also assume $R C \gg 1 / \left(2 \sigma_I \partial \omega / \partial I \right)$.  The solid lines in Fig.  $\ref{fig:Qvsftwo}$ show the results of this
procedure.  The dashed lines show the contribution to the broadening due to the
escape rate alone, while the dotted lines represent the current noise
contribution.  The parameters for the lifetime fits, $I_0 = 14.12\ \mu A, C =
3.7\ pF$ and $I_0 = 10.645\ \mu A, C = 3.7\ pF$,  agree with the parameters
obtained from Fig.  \ref{fig:wvsi}, verifying the inclusion of current noise
and escape-rate-limited coherence in the model of Eqn. \ref{eqn:withcnoise}.
We note that as the bias current approaches $I_0$ (low frequency), the escape rate term begins
to dominate the lifetime, while for lower currents (high frequency), the noise broadening
dominates.

To conclude, we have measured the resonance width of the transition between the
lowest two quantum states in a Josephson junction qubit as a function of bias
current, and found that the lifetime of the excited state falls rapidly as the
bias current $I$ approaches $I_0$.  A model including continuous dephasing from
tunneling as well as from current noise explains quantitatively the reduced
coherence time.  This ability to predict and calculate such junction behavior
is crucial to the use of junctions in quantum computers and one of the reasons
junctions are a good candidate qubit.

For designs where low-frequency current noise is not a significant issue
\cite{martinisrabi}, consideration of the above results in conjunction with
Eqn.  \ref{eqn:tunnrates} suggests the following qubit design criterion.  To
obtain at least $N_{op}$ gate operations before decoherence occurs from
tunneling alone, with each gate operation taking approximately $N_g \cdot
2\pi/\omega$ time, requires at least $N_s>\frac{5}{36}\ln (N_{op} N_p) +
\frac{5}{24} \ln (432 N_s)$ levels in the well.  For $N_{op}=10^6$ and
$N_{g}=10$, we find $N_s>4$.  In the opposite limit, where current noise
dominates, the junction must be biased at low currents during gate operations.

\begin{acknowledgments}
We acknowledge support from DOD and the Center for Superconductivity Research and thank J.M. Martinis, F. Strauch, P. Johnson, and A. Dragt for many useful discussions about this system.
\end{acknowledgments}
\bibliography{decoherence}
\end{document}